\def\year{2021}\relax
\documentclass[letterpaper]{article} % DO NOT CHANGE THIS
\usepackage{aaai21}  % DO NOT CHANGE THIS
\usepackage{times}  % DO NOT CHANGE THIS
\usepackage{helvet} % DO NOT CHANGE THIS
\usepackage{courier}  % DO NOT CHANGE THIS
\usepackage[hyphens]{url}  % DO NOT CHANGE THIS
\usepackage{graphicx} % DO NOT CHANGE THIS
\usepackage{natbib}  % DO NOT CHANGE THIS AND DO NOT ADD ANY OPTIONS TO IT
\usepackage{caption} % DO NOT CHANGE THIS AND DO NOT ADD ANY OPTIONS TO IT
\usepackage{bbm}
\usepackage[ruled,vlined]{algorithm2e}
\usepackage{algorithmic}
\usepackage{colortbl}
 
\newcommand{\change}[1]{#1}

\newcommand{\hide}[1]{}
\newcommand{\cut}[1]{}

\title{Deception Detection in Group Video Conversations\\ using Dynamic Interaction Networks}

\author{\Large Srijan Kumar,\textsuperscript{\rm 1} Chongyang Bai,\textsuperscript{\rm 2} 
V.S. Subrahmanian,\textsuperscript{\rm 2} 
Jure Leskovec\textsuperscript{\rm 3}
}
\affiliations{\\
\textsuperscript{\rm 1}Georgia Institute of Technology\\
\textsuperscript{\rm 2}Dartmouth College\\ \textsuperscript{\rm 3}Stanford University\\ 
srijan@gatech.edu, chongyang.bai.gr@dartmouth.edu, vs@dartmouth.edu, jure@cs.stanford.edu
}
 
\begin{document}

\maketitle

\begin{abstract}
Detecting groups of people who are jointly deceptive in video conversations is crucial in settings such as meetings, sales pitches, and negotiations. Past work on deception in videos focuses on detecting a single deceiver and uses facial or visual features only. In this paper, we propose the concept of Face-to-Face Dynamic Interaction Networks (FFDINs) to model the interpersonal interactions within a group of people. The use of FFDINs enables us to leverage network relations in detecting group deception in video conversations for the first time. We use a dataset of 185 videos from a deception-based game called Resistance. We first characterize the behavior of individual, pairs, and groups of deceptive participants and compare them to non-deceptive participants. Our analysis reveals that pairs of deceivers tend to avoid mutual interaction and focus their attention on non-deceivers. In contrast, non-deceivers interact with everyone equally. We propose Negative Dynamic Interaction Networks to capture the notion of missing interactions. We create the DeceptionRank algorithm to detect deceivers from NDINs extracted from videos that are just one minute long. We show that our method outperforms recent state-of-the-art computer vision, graph embedding, and ensemble methods by at least 20.9\% AUROC in identifying deception from videos.
\end{abstract}

\def\FFIN{{\textsf{FFDIN}}}
\def\FFINs{{\textsf{FFDIN}s}}

\section{Introduction}
% \vspace{-0.2em}

Web-based face-to-face video conversations have become a pervasive mode of work and communication throughout the world, especially since the COVID-19 pandemic. Important tasks, such as interviews, negotiations, deals, and meetings, are all happening through video call platforms such as Microsoft Teams, Google Meet, Facebook Messenger, Zoom and Skype. Furthermore, video content have become a central theme in social media and video conversations have also become an integral part of social media platforms, including on Facebook, WhatsApp, and SnapChat. Deception and disinformation in all these settings can be disruptive, counter-productive, and dangerous. 

The problem of accurately and quickly identifying whether a group of people is being deceptive is crucial in many settings. 
Specifically, consider the scenario where a group of deceivers work together to fool a group of unsuspecting users, but the latter do not know who the deceivers are. 
This occurs in practice, for instance, when defectors are present in security teams, when liars are present in sales teams, and when people from competing firms infiltrate an organization.

While there has been significant research in identifying individual deceivers in real-world face-to-face interactions~\cite{wu2018deception,ding2019face,gogate2017deep}, little is known about how groups of deceivers work together in the online setting. 
Current deception research is largely limited to analysing the audio-visual behavior of a single deceiver using voice signatures, gestures, facial expressions, and body language. 
In contrast, research on social media analytics has extensively studied the impact of individual and team of deceivers~\cite{kumar2017army,kumar2015vews,kumar2019predicting,addawood2019linguistic,wu2017adaptive,keller2017manipulate}, but those findings do not translate to the case of video-based face-to-face group deception. 

The behavioral characteristics when multiple deceivers operate simultaneously are drastically different from the case of a single deceiver.  This is primarily because the behavior of one deceiver can influence the behavior of the other deceivers. For instance, when one deceiver lies to a potential target, other partners of the deceiver may show certain facial reactions which might be leveraged to predict that deception is going on.
Moreover, multiple simultaneous deceivers, each deceiving a little bit, may be more successful in deceiving victims than a single deceiver alone. 
There is little work on studying the behavioral patterns of groups of deceivers, which is a gap that we bridge.

\begin{figure*}[!t]
\centering
        \includegraphics[width=0.9\textwidth]{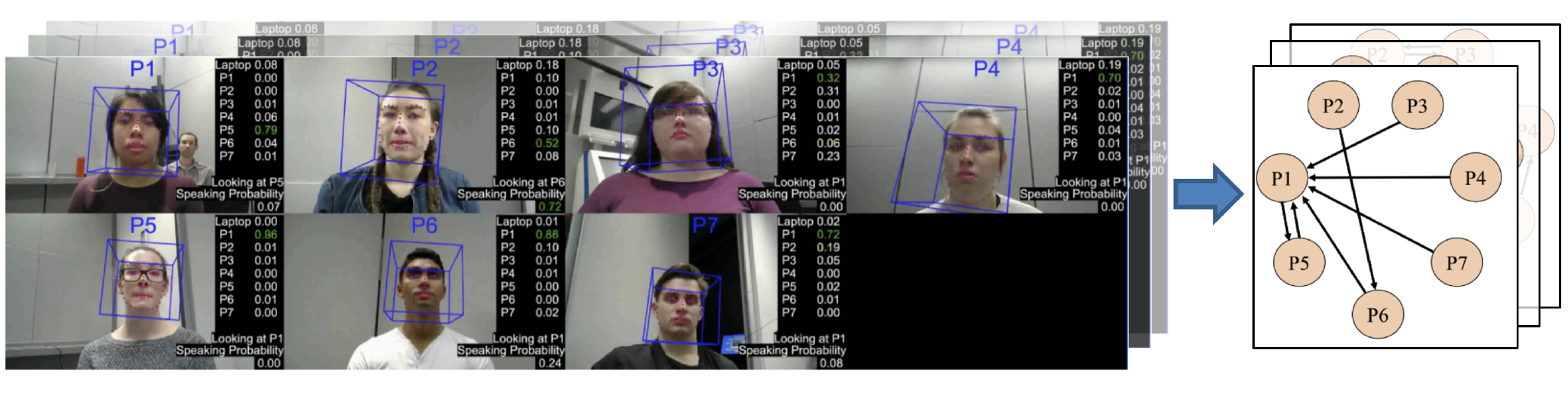}
    \caption{Given a group video conversation (left), we extract face-to-face dynamic interaction networks (right) representing the instantaneous interactions between participants. Participants are nodes and interactions are edges in the network. In this work, dynamic interaction networks are used to characterize and detect deception.
    \label{fig:example}}
\end{figure*}

%\begin{figure*}[!t]
%\centering
% \includegraphics[width=0.8\textwidth]{fig/figure-example-v6.png}
%        \includegraphics[trim={0cm 6cm 0cm 6cm},clip,width=0.8\textwidth]{fig/figure_mafia_example2.pdf}
%        \includegraphics[trim=0 -4cm 0 0cm, width=0.18\textwidth]{fig/network_mafia_example4.png}
%    \caption{Given a group video conversation (left), we extract face-to-face dynamic interaction networks (right) representing the instantaneous interactions between participants. Participants are nodes and interactions are edges in the network. In this work, dynamic interaction networks are used to characterize and detect deception.
%    \label{fig:example}}
%\end{figure*}

We conduct the first network-based study of group deception in face-to-face discussions. We analyse the verbal behavior, non-verbal behavior, and inter-personal interaction behavior when multiple deceivers are present within a group of people. We elicit deceptive behavior in the form of a multi-person face-to-face game called \textit{Resistance}\footnote{Many variants of Resistance such as Mafia and Werewolf are played around the world by thousands of people.}.
Resistance is a social role-playing card-based party game, where a small group (deceivers) tries to disrupt the larger group (non-deceivers) working together.
In this study, we use a dataset of 26 Resistance games, each 38-minutes long on average~\cite{bai19vfoa}.

We propose the concept of a Face-to-Face Dynamic Interaction Network (FFDIN for short) which captures instantaneous interactions between the participants. Participants are nodes and interactions are edges in the network. FFDINs include both verbal (who talks to whom) and non-verbal (who looks at whom) interactions. One FFDIN is extracted per second of the video. An example is shown in Figure~\ref{fig:example}. As discussions unfold over the course of the game, FFDINs evolve rapidly over time and their dynamics can provide valuable clues for detecting deception.
We use the dynamic FFDINs extracted from the videos of the Resistance games in this work.
Even though 26 games does not plenty, 
our dataset consists of $59,762$ FFDINs in total.

We conduct a series of analysis on these networks which reveal novel behaviors of deceivers, extending the research done in social sciences~\cite{driskell2012social,baccarani2015effective,vrij2008detecting}. In particular, we find that deceivers who are less engaged (as measured by the number of participants they interact with, how often they speak, and who listens to them) are more likely to lose the game. 
On the other hand, deceivers successfully deceive others when they are as engaged in the game as non-deceivers, thus adeptly camouflaging themselves.
Across all games, we also found that deceivers interact significantly more with non-deceivers than with other deceivers, echoing previous findings by \change{\cite{driskell2012social}}. In contrast, as non-deceivers do not know the identity of other participants, they interact equally with everyone. 

We introduce the notion of Negative Dynamic Interaction Networks (or NDINs) that captures when two participants \textit{avoid} interacting with one another. We then create an algorithm called DeceptionRank  that can detect deceivers even from very short (one minute) video snippets. 
DeceptionRank derives NDINs from the original interaction networks --- two nodes are linked with an edge if their corresponding participants do not interact. Our method initializes the prior deception scores of each node through a novel process which normalizes and aggregates the verbal and non-verbal behaviors of nodes. It then iteratively runs PageRank on and aggregates scores from the set of negative dynamic interaction networks. This generates a deception score for each node.
We show that DeceptionRank outperforms the state-of-the-art computer vision, graph embedding, and ensemble methods by over 20.9\% AUROC in detecting deceivers. 
Moreover, we show that DeceptionRank is consistently the best performing method across different lengths of video segments and regardless of the final outcome of the game.

The dynamic networks dataset along with ground truth of deception are available at:
\url{https://snap.stanford.edu/data/comm-f2f-Resistance.html}.

\section{Dataset Description}
% \vspace{-0.1em}
Here we create face-to-face dynamic verbal and non-verbal interaction networks by extending the work by \citet{bai19vfoa}.
In \citet{bai19vfoa}, face-to-face interactions are extracted from videos of a group of participants playing Resistance game.\footnote{\url{https://en.wikipedia.org/wiki/The_Resistance_(game)}} The extraction algorithm is a collective classification algorithm that leverages computer vision techniques for eye gaze and head pose extraction. %Please refer to the paper for complete details of how the interactions are extracted. 
Each game has 5--8 participants, out of which a subset are assigned the roles of being deceivers (others are non-deceivers). A participant has the same role throughout the game. The deceivers know who the other deceivers are, but the non-deceivers do not know the roles of any other participant. One participant is part of exactly one game. In total, the dataset has 26 game and 185 participants.

The game has multiple rounds. Each round starts with a free-flowing discussion in which players discuss who the possible deceivers might be. Players cast votes at the end of each round. 
To win the game, the non-deceivers must collectively identify the deceivers as early as possible --- but as they do not know who the deceivers and non-deceivers are, they must identify who is lying. 
The dominant winning strategy of the non-deceivers is to be truthful and that of deceivers is to lie and pretend that they are non-deceivers. 
In our dataset, deceivers win 14 out of 26 games (or 54\%). Hence the data is reasonably balanced.
We use  ``DW''  to label the games that the deceivers win and ``DL'' to identify games that the deceivers lose.

{ 
\begin{table}[t]
% \footnotesize
\centering

% \vspace{-6mm}}
\vspace{1em}
\resizebox{\columnwidth}{!}{
\begin{tabular}{c | c | c }
\hline
\textbf{Property} & \textbf{Total} & \textbf{Per game} \\
\hline
Number of network timeseries & 26 & 1 \\
Temporal length (in seconds) & 59,762 & 2,299 \\
Number of games won by deceivers & 14 & 54\% \\
Number of nodes & 185 & 5--8\\ 
Number of look-at edges & 689,501 & 26,519\\
Number of speak-to edges  & 26,556 & 1,021\\
Number of listen-to edges  & 25,798 & 992\\
\hline
\end{tabular}
}
\caption{\label{tab:dataset} Statistics of the Resistance dataset.}
\end{table}
}

\emph{Face-to-Face Dynamic Interaction Networks (FFDINs).}
We create a dataset of FFDINs. Each game is represented as a sequence of interaction networks, with one network snapshot per second. 
Nodes in an FFDIN represent participants in the corresponding game. 
Each node has a binary attribute representing its role, i.e., deceiver or non-deceiver (a participant's role does not change during the game).
An edge represents the interaction between a pair of participants during the corresponding second -- we will discuss the types of edges considered shortly. 
The resulting FFDINs have highly dynamic edges because of the free-form discussion and interaction changes over time.
All edges are directed and weighted --- the weight indicates the strength or probability of the interaction. 
An example is shown in Figure~\ref{fig:example}.
In total, there are 26 games (network sequences), 185 participants, and 996 minutes of recordings. 
Table~\ref{tab:dataset} shows the statistics of the Resistance networks.
We create three types of networks from the video of each game:
\begin{itemize}
\item \textit{Look-At FFDINs} $N_G^t=(V,E_G^t)$ captures non-verbal interactions between participants. The edges at time $t$ represent who-is-looking-at-who during the time duration $t$. The edge weight $E_G^t(u,v)$ is the probability of participant $u$ looking at participant $v$ at time $t$.
\item \textit{Speak-To FFDIN} $N_S^t=(V,E_S^t)$ captures verbal interactions between participants. The edges represent who speakers are looking at while speaking. At any time point, the edges emanate from speaker nodes. Edge weights represent the probabilities of speakers looking at targets. 
\item \textit{Listen-To FFDIN} $N_L^t=(V,E_L^t)$ shows who listens to the speaker. The edges are incoming weighted edges directed towards the speaker node at each point in time.
\end{itemize}

\subsection{Ethical and IRB Considerations}
An extensive set of IRB approvals were obtained by the authors of \citet{bai19vfoa} to collect the data. IRB review was conducted at the institutions where the data was collected as well as the IRB of the project sponsor. The participants gave permission to the research team of \citet{bai19vfoa} to record and analyze their videos. After the networks are extracted from \citet{bai19vfoa}, all personally identifiable information (PII) is stripped and original videos are not used further. 

Our data and networks are derived using the interaction information provided as output by \citet{bai19vfoa}, which has no PII. As a result, the FFDINs created in this work do not contain any PII as well. The dynamic networks dataset we release in this work does not have any PII as well.

\section{Characterizing Deceptive Behavior}
\label{sec:analysis}
The goal of this section is to answer two important questions:
(i) what are the behavioral characteristics that separate deceivers from non-deceivers? and (ii) what are the factors that distinguish \emph{successful} deceivers from unsuccessful ones?
We answer these questions through three research questions.

\subsection{RQ1: Do deceivers and non-deceivers have distinct gaze patterns?}
% \vspace{-0.3em}
There is an asymmetry between the knowledge that deceivers and non-deceivers have. In the game, deceivers know who the deceivers and non-deceivers are. In contrast, a non-deceiver only knows her own role and nothing whatsoever about the other participants. 
Since deceivers know the roles of all participants, a natural question to ask is whether they focus their attention on specific participants? 
If yes, how does this focus affect their success in deceiving others? 
This is a key question as prior social science research has shown that frequent/rapid gaze change is linked to low confidence~\cite{rayner1998eye} and higher likelihood of both deception~\cite{pak2013eye} and anxiety~\cite{dinges2005optical,laretzaki2011threat}, which may be exhibited by users in certain roles. 
For instance, prior research has found that deceivers are more anxious than non-deceivers~\cite{strofer2016catching}.

We analyse the behavior of deceivers in the Look-At networks. 
A participant $u$'s ``looking'' behavior can be represented as a sequence of consistent gaze periods [$P_{u1}, P_{u2}, \ldots P_{un}$]. 
A period $P_{uk}$ is a continuous time interval $[P_{uk}^0, P_{uk}^1]$ with a single gaze target $T_{uk}$, i.e., the recipient of $u$'s highest weight outgoing edge at every time step in the interval. 
Suppose we  use $D_{uk}$ to denote the duration of participant $u$'s $k^{th}$ period $P_{uk}$. % of participant  the time duration for which the target remains $u$'s highest weighted outgoing edge. 
Thus, the duration sequence for $u$'s gaze behavior is represented as [$D_{u1}, D_{u2}, \ldots D_{uk}$], where $D_{u1} +  D_{u2} + \ldots + D_{uk} = T$, i.e. 
$D_{u1}, D_{u2}, \ldots D_{uk}$ partition the time interval $T$.

\textbf{Gaze Entropy.}
We calculate the entropy of $u$'s gaze behavior in the game as the entropy of the set \{$\frac{D_{u1}}{T}$, $\frac{D_{u2}}{T}$, \ldots $\frac{D_{uk}}{T}$\}. Specifically, $ H_u = -\sum_{i=1}^k \frac{D_{ui}}{T}\log(\frac{D_{ui}}{T})$.
A high entropy value $H_u$ means that $u$ changes his/her focus of attention very frequently, indicating more engagement in the game. 
Conversely, a low  $H_u$ means longer periods of gaze towards the same participant, indicating lower engagement with the rest of the group. 

All scores are normalized per game by subtracting the mean score of all participants in the game. 
Thus, after normalization, a positive (negative) gaze entropy of a participant $p$ means that $p$ shifts her gaze more (less) often than average. 
To compare the overall behavior of deceivers and non-deceivers, we average the normalized entropy score of all deceivers' entropy across all games --- and likewise do the same with non-deceivers. Furthermore, since the behavior of participants can vary dramatically based on the game's outcome (i.e., whether deceivers win or lose), we aggregate the scores for DW (Deceivers Win) vs DL (Deceivers Lose) games separately.

\textbf{Gaze Reciprocity.}
We define the reciprocity of $u$'s gaze in the $k^{th}$ period $P_{uk}$ as the average looking-at probability of $u$'s target $T_{uk}$ towards $u$ during the same time period. 
Specifically, $R_{uk} = \frac{\sum_{t = P_{uk}^0}^{P_{uk}^1} E_G^t(T_{uk}, u)}{D_{uk}}$.
A high reciprocity means that $u$'s targets pay attention to $u$, while a lower reciprocity indicates that $u$'s targets ignore $u$'s gaze.
The average reciprocity of $u$ in the game is the average reciprocity across all its periods, weighted by the duration of the period.
We normalize and aggregate reciprocity scores to zero-mean as we did with entropy.

\begin{figure}
\centering
        \includegraphics[width=0.45\columnwidth]{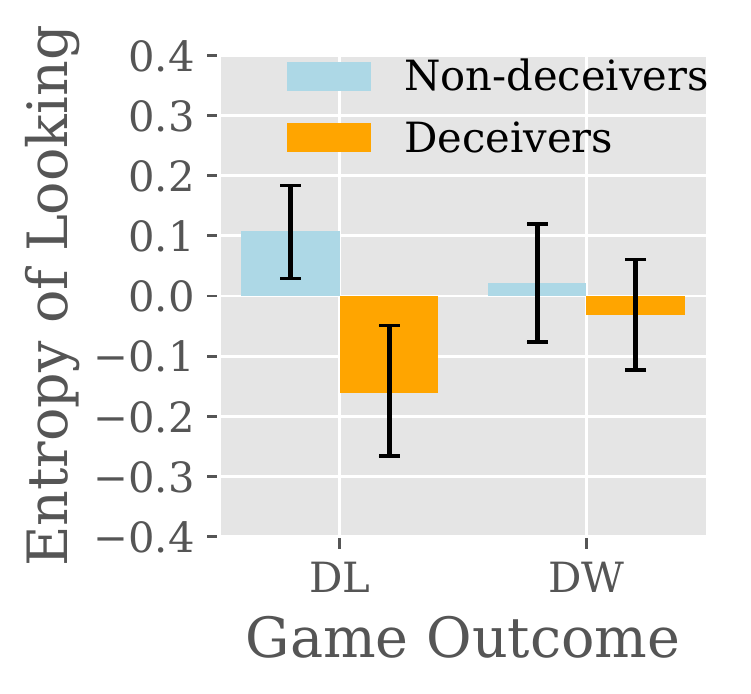}
        \includegraphics[width=0.45\columnwidth]{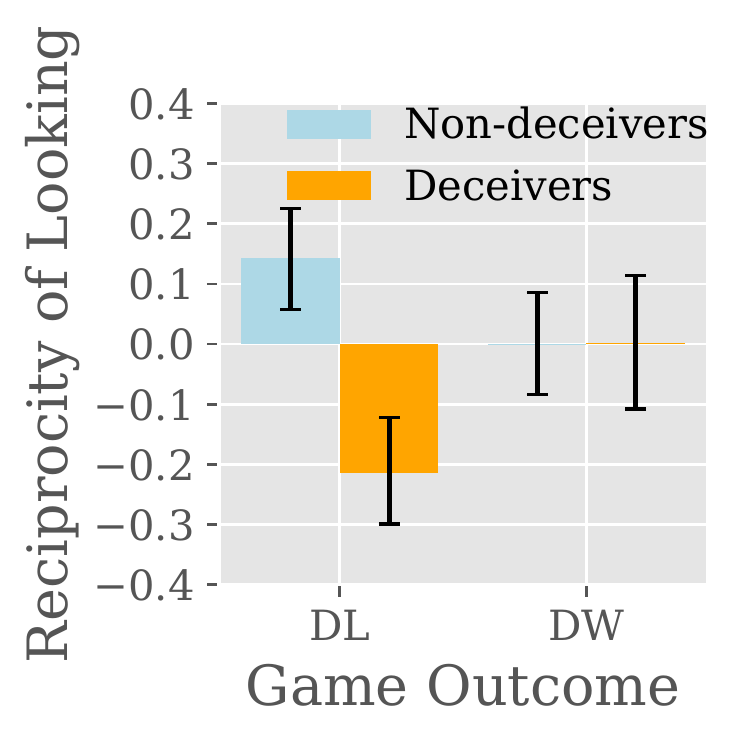} 
    % \vspace{-1em}
    \caption{\textit{Non-engaging deceivers are unsuccessful in deceiving others.}
    (Left) Deceivers have a lower entropy of looking at others compared to non-deceivers in DL (Deceivers Lose) games, while this difference does not exist in DW (Deceivers Win) games. 
    (Right) Deceivers get less attention, measured in terms of reciprocity, in DL games. 
    In both plots, we observe that deceivers and non-deceivers have similar scores in DW games, as deceivers successfully camouflage themselves.
    % \vspace{-4mm}
    \label{fig:look-at-pattern}
    }
\end{figure}

\textbf{Findings.}
Figure~\ref{fig:look-at-pattern}(left) and (right) respectively 
compare the gaze entropy of participants and their gaze reciprocity. The figures report the mean scores across participants and the 95\% confidence interval of the score distribution.
An independent two-sample t-test is used to compare distributions throughout the paper.
We observe that the behavior of deceivers depends heavily on the  outcome of a game.
    
\textbf{Finding (F1): In DW games, deceivers and Non-Deceivers look at similar numbers of speakers and are looked at to similar extents by other participants.}
Deceivers and non-deceivers have similar entropy and reciprocity scores (both $p > 0.05$) in DW games.
Thus, deceivers win when they successfully camouflage themselves by imitating the nonverbal behavior of non-deceivers.

\textbf{Finding (F2): In DL games, deceivers look at fewer participants than non-deceivers and are also looked at less by other participants.} 
Both gaze entropy and gaze reciprocity are significantly lower for deceivers in DL games ($p < 0.001$), showing that they have a steadier gaze and receive less attention compared to non-deceivers.
This indicates that deceivers are easily identified and lose when they are less engaging as compared to the rest of the participants.

%%%%%%%%%%%%%%%%%%%%%%%%%%%%%%%%%%%%%%%%%%%%%%%%%%%%%%%%%%%%%%%%%%%%%%

\subsection{RQ2: Does the verbal behavior of group of deceivers differ from non-deceivers?} 
% \vspace{-0.2em}
The way in which people speak has previously been shown to indicate deception ~\cite{baccarani2015effective,beslin2004leaders}. 
However, verbal characteristics of a coordinated group of deceivers is less well known. We now compare the speaking patterns of deceivers and non-deceivers. We use the Speak-To FFDIN network for the analysis. 
As before, we compare the differences partitioned by the final game outcome. 

\paragraph{Finding (F3): Deceivers speak less than non-deceivers.} We established this by first computing the fraction of time slices in which a participant $u$ speaks.
Both speaking excessively and being anomalously quiet have previously been noted to be indicators of deception~\cite{wiseman201059,vrij2008detecting}.

\begin{figure}
\centering
\includegraphics[width=0.55\columnwidth]{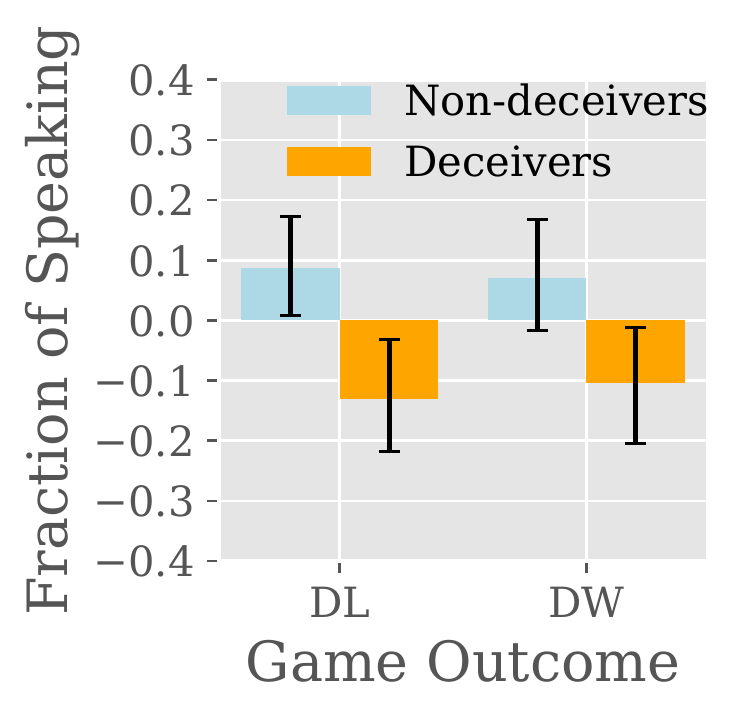}
% \vspace{-1em}
\caption{\textit{Deceivers tend to speak less than non-deceivers.} This difference is more pronounced in DL games.  \label{fig:speaking}}
% \vspace{-1em}
\end{figure}

Our experiments show that deceivers speak less than non-deceivers regardless of the game outcome.  \change{As shown in Figure~\ref{fig:speaking}}, 
the differences are statistically significant both in DW games ($0.08$ vs $-0.13, p < 0.05$) and in DL games ($0.07$ vs $-0.10, p < 0.001$). 
This shows that non-deceivers are more vocal in all games.

\vspace{-2mm}
\paragraph{Finding (F4): In DL games, deceivers get less attention while speaking than non-deceivers.} We can infer the attention a speaker $u$ is getting based on how many other participants are looking at $u$ while $u$ is speaking. We define the average attention that a participant $u$ gets as the average weighted in-degree of $u$ in the Listen-To FFDINs: $\frac{1}{T}\sum_t\sum_{(v,u) \in E^t_L}E^t_L(v,u)$, where $T$ is the number of networks in which $u$ is a speaker.

\begin{figure}
\centering
        \includegraphics[width=0.49\columnwidth]{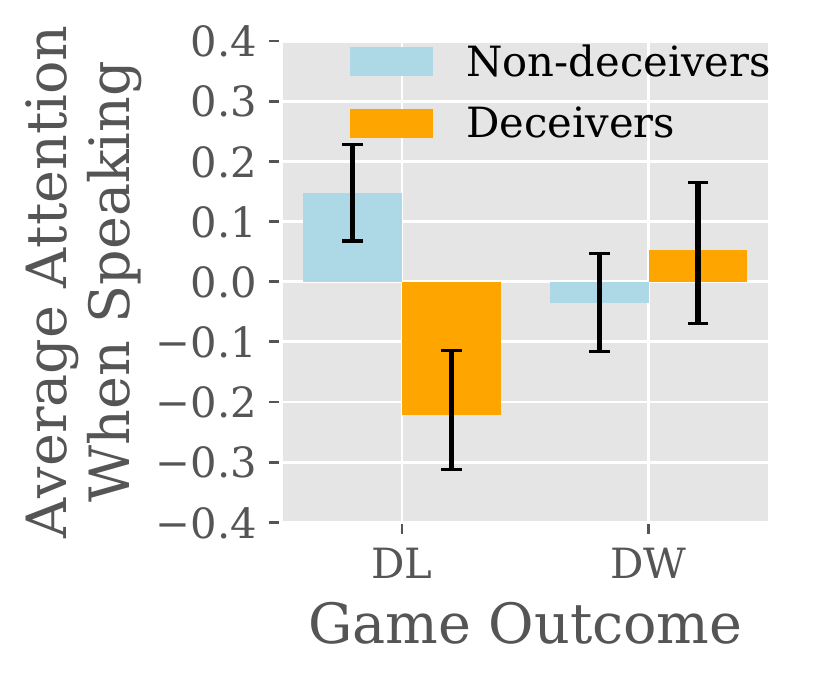} 
        \includegraphics[width=0.45\columnwidth]{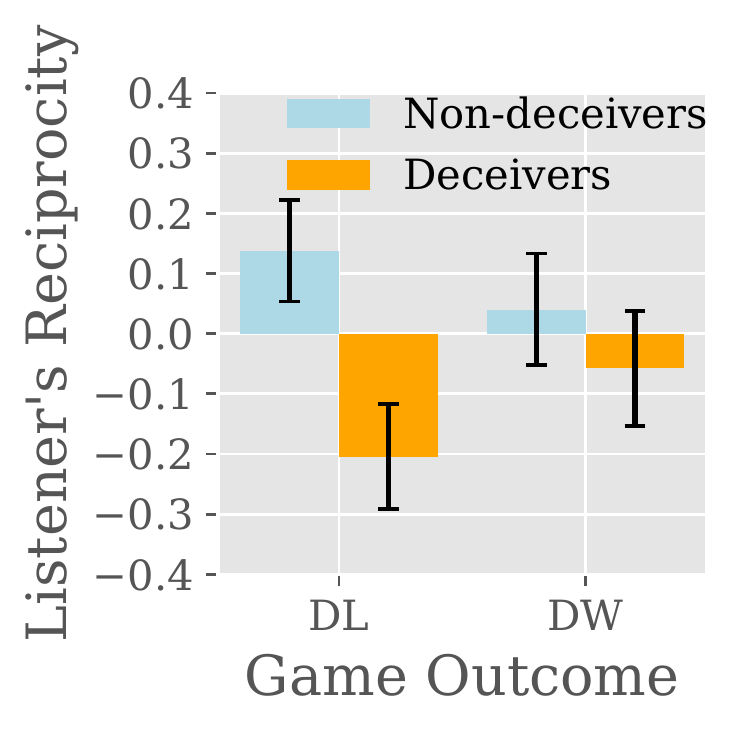} 
        % \vspace{-1em}
    \caption{
    \textit{Deceivers get less attention in the games they lose.} 
    (Left) We observe that when deceivers speak in DL games, other participants have lower likelihood of looking at the deceiver as compared to when a non-deceiver is speaking.
    (Right) Similarly, the target of speakers is less likely to look back at deceivers than at non-deceivers in DL games. 
    These differences are not present in DW games as deceivers are equally engaged and central to discussions. 
    % \vspace{-4mm}
    \label{fig:listening}
    }
\end{figure}

Figure~\ref{fig:listening} (left) shows that in DL games, less attention is paid to deceivers when they speak, compared to non-deceivers. However, the story is different in DW games--- players in both roles receive a similar amount of attention.

\vspace{-2mm}
\paragraph{Finding (F5): In DL games, deceiver speakers are reciprocated less than non-deceivers.}
We also looked at the gaze behavior of the person who is being spoken to. How often do they pay attention to the speaker? Specifically, when person $u$ talks to $v$, does $v$ look back at $u$? This is a sign of trust and respect~\cite{ellsberg2010power,derber2000pursuit}.

We define the reciprocity of $u$'s target $T_{ut}$ at time $t$ as the edge weight from $T_{ut}$ to $u$ at the same time. We calculated the average reciprocity of participant $u$ in the entire time period $t \in [1,T]$ as $\frac{1}{T}\sum_{t=1}^T E_L^t(T_{ut}, u)$.
We compared the average reciprocity of deceivers and non-deceivers in Figure~\ref{fig:listening} (right).
We see that in DL games, deceivers are not as frequently reciprocated as non-deceivers. This suggests that in DL games,  other participants pay less attention to and trust deceivers less.
 This, however, is not the case in DW games where both deceivers and non-deceivers are given equal attention by listeners.

Summarizing, we find that non-deceivers are highly vocal and more active and central compared to deceivers in DL games. 
This is not the case in DW games, where the engagement and importance of deceivers and non-deceivers is equivalent. This shows that deceivers are successful in deceiving others when they are as engaging as the non-deceivers in the game and camouflage their behavior well.

\begin{figure}[t]
\centering
        \includegraphics[width=\columnwidth]{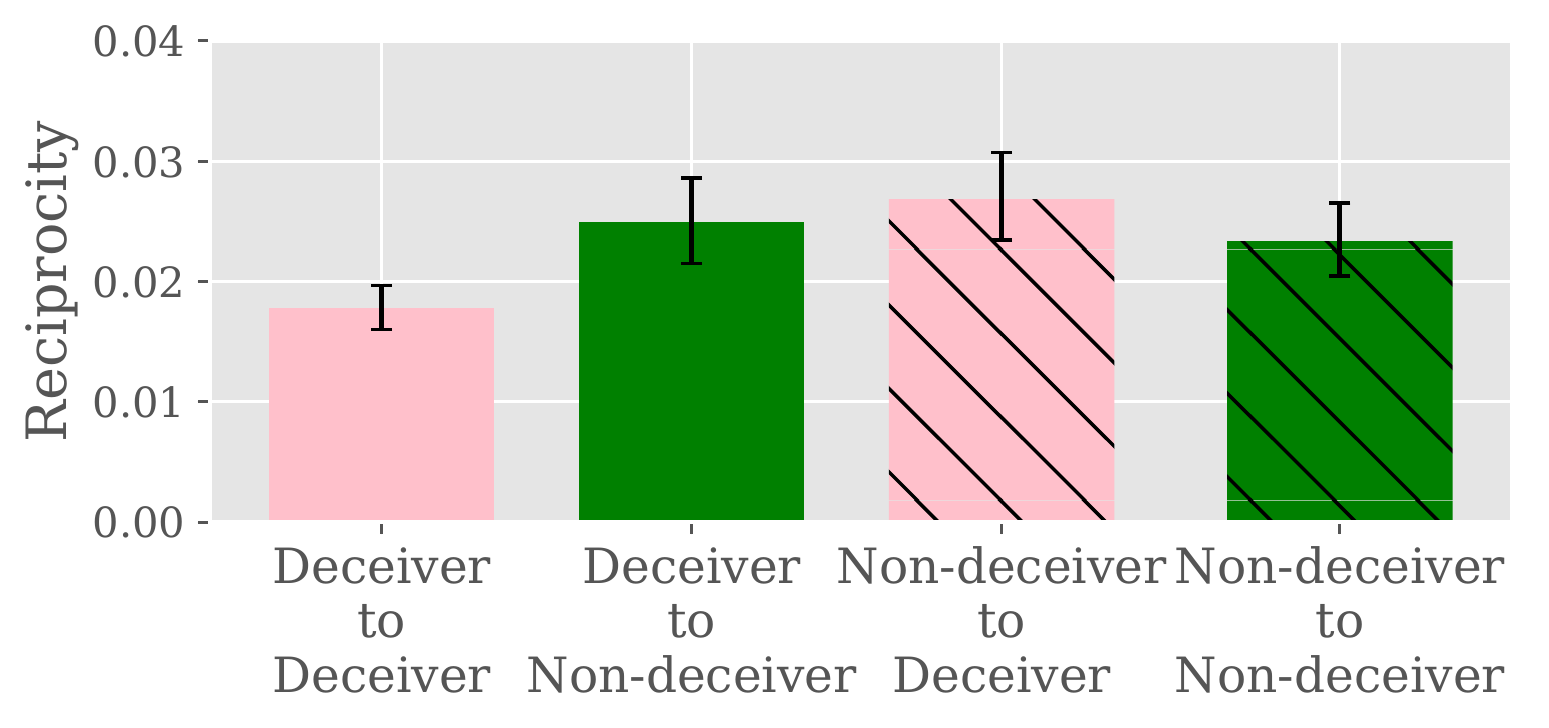}
        
        \includegraphics[width=\columnwidth]{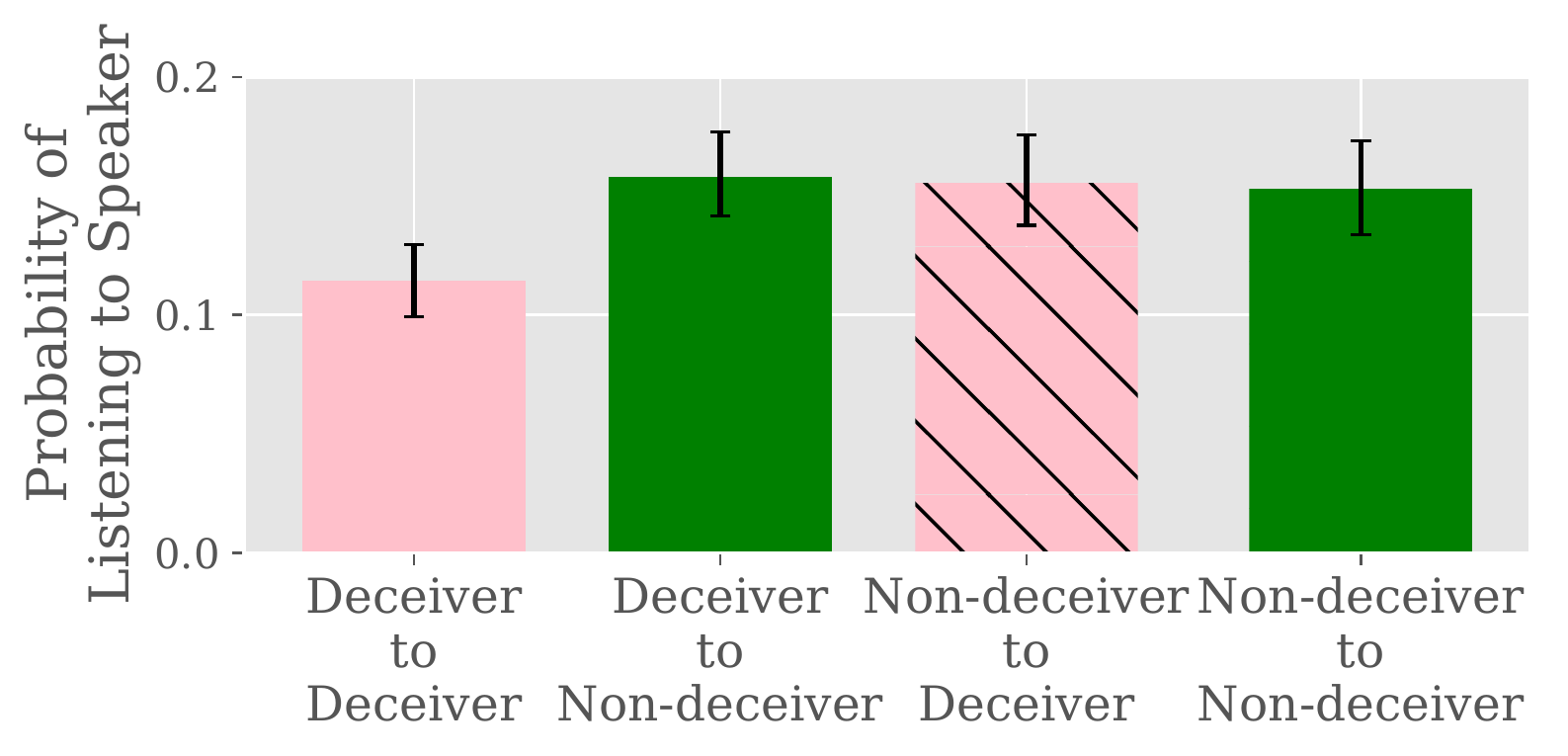}
        
        \includegraphics[width=\columnwidth]{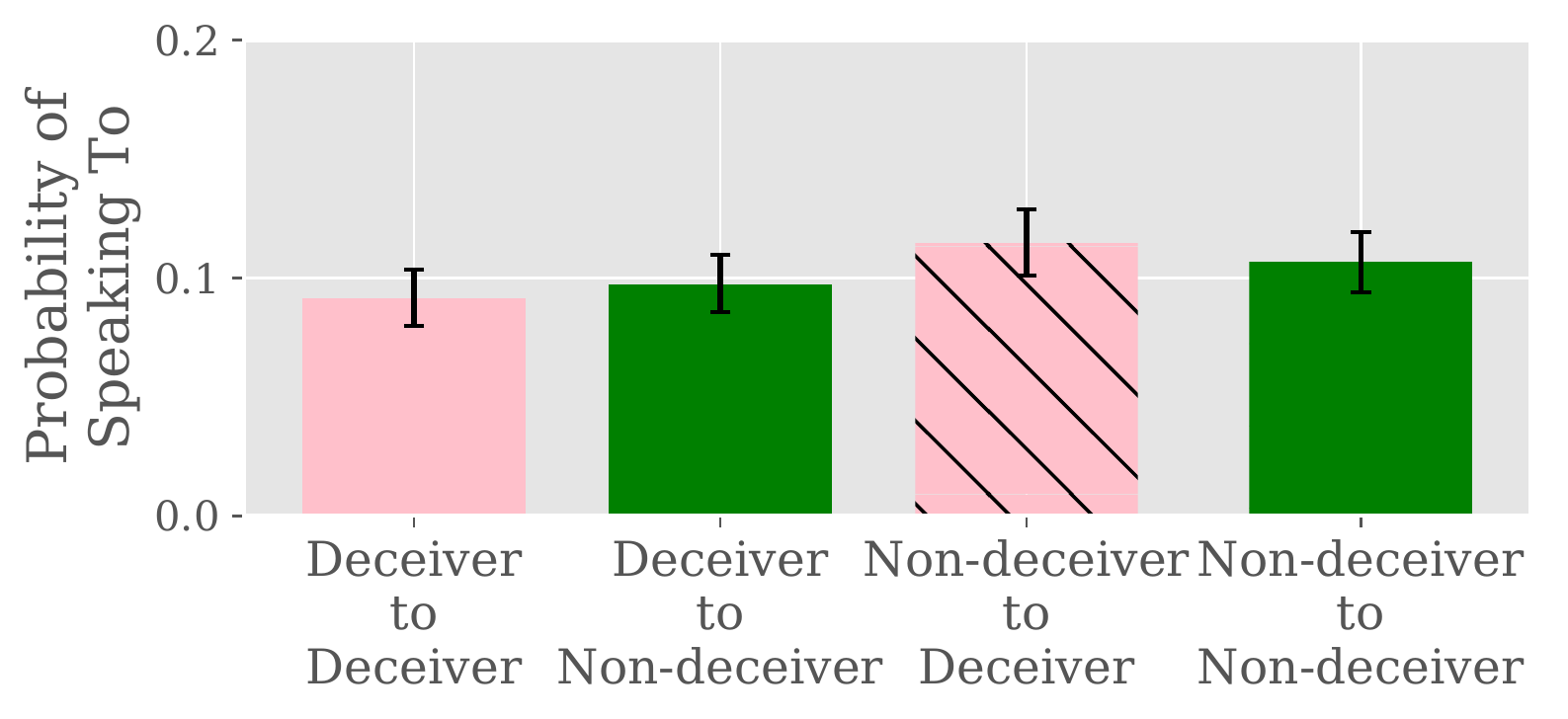}
        % \vspace{-1em}
    \caption{
    \textit{Deceivers avoid non-verbal interactions with other deceivers.}
    (Top) Deceivers have lower reciprocity of looking at other deceivers than at non-deceivers. 
    Non-deceivers have similar reciprocity for both. 
    (Middle) Deceivers listen less to other deceivers than to non-deceivers. Non-deceivers have similar values for both.
    (Bottom) All pairs of participants have similar probability of speaking to one another. 
    \label{fig:pairwise-interact}}
\end{figure}

\subsection{RQ3: Do deceivers interact differently with other deceivers vs non-deceivers?}
% \vspace{-0.3em}
Since deceivers know the role of all participants in the game, do they focus their attention on specific individuals? Past survey based social science studies ~\cite{driskell2012social} conclude that deceivers are unlikely to respond to each other. We develop competing hypothesis about this. 
The first hypothesis is that the deceivers interact more with deceivers in order to cooperate and deceive other participants.
The alternate hypothesis states that deceivers interact less with each other in order to avoid being identified by non-deceivers.

To test these hypotheses, we compare the pairwise interactions between participants, grouped by their roles:  deceivers vs. deceivers, deceivers vs. non-deceivers, non-deceivers vs. deceivers, and non-deceivers vs. non-deceivers. 
Figure~\ref{fig:pairwise-interact} compares the average reciprocity of looking, average talk-to probability, and average listen-to probability for all pairs of roles. 
We aggregate the properties across all games to measure role-specific behavior regardless of the game's outcome. 
As earlier, we report the mean and the 95\% confidence intervals for all properties. 
We make the following three observations by analysing both verbal and non-verbal behavior of participants.

\textbf{Finding (F6): Deceivers look less at other deceivers.}
First, Figure~\ref{fig:pairwise-interact} (top) shows that non-deceivers spend similar amounts of time looking at deceivers and non-deceivers ($p > 0.05$). However, the looking behavior of deceivers is strikingly distinct---deceivers look less at other deceivers than at non-deceivers ($p < 0.001$).
This has several important implications. 
Since deceivers know the identity of non-deceivers, deceivers spend more time observing non-deceivers (and less time observing their fellow deceivers). 
Deceivers may also interact less with other deceivers to avoid `guilt-by-association', i.e., getting caught in case the other deceivers are identified. 

In addition, Figure~\ref{fig:pairwise-interact} (middle) shows that non-deceivers have a similar probability of listening to both deceptive and non-deceptive speakers ($p=0.76$). However, this is not the case for deceiver listeners. Deceivers have a lower probability of listening to other deceivers as compared to non-deceivers ($p < 0.05)$. 
This is possibly in order to avoid being suspected of supporting the deceiver. 

Finally, Figure~\ref{fig:pairwise-interact} (bottom) compares the verbal behavior (as opposed to non-verbal behavior in the previous two paragraphs). 
Surprisingly, we find that the verbal behavior between all pairs of participants is similar. 
Since non-deceivers do not know the roles of other participants, they speak equally to non-deceivers and deceivers ($p=0.48$), as expected. 
However, it is surprising that deceivers spend equal time talking to both deceivers and non-deceivers ($p=0.39$). 
This is in stark contrast to the previous two non-verbal findings. 
This shows that deceivers consciously adapt their verbal behavior to mimic non-deceivers, but not their non-verbal behavior. 
Since verbal behavior is noticed by everyone else, deceivers consciously do not exhibit any bias in verbal interaction with other participants to avoid getting caught.

Thus, RQ3 shows that deceivers successfully camouflage their verbal (speaking) behavior, while they are unable to camouflage their non-verbal (looking and listening) behavior.
Altogether, deceivers avoid non-verbal interactions with other deceivers.

\section{Network Model for Deception Prediction}
In this section, we present DeceptionRank, our PageRank based model that examines FFDINs in order to predict whether a given participant is deceptive or not.
Automated detection of deceivers is a challenging task as this is the goal of the non-deceivers. DeceptionRank tries to do this with short duration of videos. In contrast, human participants try to do this throughout the game, which are 38 minutes long, on average, but are still unsuccessful in almost half the games.

\subsection{DeceptionRank on Negative Dynamic Interaction Networks}

DeceptionRank is built on our findings from the previous section 
that deceivers avoid non-verbal interactions with other deceivers, while non-deceivers do not exhibit this bias. 
There are four main steps of DeceptionRank: (i) building the network, (ii) initializing node deception priors, (iii) applying network algorithm to obtain node deception scores, and (iv) training a deception classifier.

\textbf{Building negative dynamic interaction networks.}
In order to bring this ``non-interaction'' to the fore, we generate negative interaction networks that capture the pairwise ``lack of interactions''. The edges in the negative interaction network connect nodes which avoid interacting with one another. Given a FFDIN $N^t=(V,E^t)$ at time $t$, where $E^t(u,v)=w_{u,v,t}$,  $\forall u,v \in V$, the associated negative interaction network NDIN is given by $N^{t-}=(V,E^{t-})$, where $E^{t-}(u,v)=1-w_{u,v,t}$. Note that $E^{t-}(u,v) = 1$ when there is no edge from $u$ to $v$ at time $t$ in the interaction network, i.e., when $u$ does not interact with $v$.

\textbf{Initializing node deception priors.}
First we need to initialize every node's prior probability of being a deceiver. % in the network for belief propagation.
We introduce a novel technique for initialization based on every node's verbal and non-verbal features compared to the features of all the nodes in the network. 
Given a set of feature values $\{x_{1u},...,x_{Fu}\}$ for the $F$ features of a node $u$, we aim to combine them into an initial deception score $S(u) \in [0, 1]$, which we describe next. 

Based on our analysis in the previous section, we build the priors using the following four features that best distinguish between deceivers and non-deceivers: (a) fraction of speaking ($FS_u$), (b) average entropy of looking ($H_u$), (c) average in-degree ($E_{G,u}$), and (d) average in-degree while speaking ($E_{L,u}$).
Since the feature distributions can vary, we first normalize each feature $f (f \in \{FS_u, H_u, E_{G,u}, E_{L,u}\})$ by linearly scaling it between 0 and 1, corresponding to the minimum and maximum values of the feature. 
Then we subtract each feature $f$ from 1 because deceivers have lower scores of $f$ than non-deceivers, so $1-f$ ensures that the deceivers tend to have higher initial scores. 
Finally, we average each node's four property scores to get its prior score $S(u)$ for node $u$:
\begin{equation}
% S_u = 1 - \frac{1}{4}(fs_u+H_u+ E_{G,u}+E_{L,u})
S(u) = 1 - \frac{FS_u+H_u+ E_{G,u}+E_{L,u}}{4}
\end{equation}
This score is used to initialize node priors in the first iteration of our dynamic network algorithm --- a higher score would indicate higher prior probability of the node being deceptive.

\textbf{Obtaining node deception scores from negative networks.}
To predict whether a participant is a deceiver or not, we extend the PageRank algorithm~\cite{page1999pagerank}. By default, PageRank is applicable on static networks, thus, we extend it to apply to dynamic negative interaction network sequences. 
The method is shown in  Algorithm~\ref{alg:bp}.
The overall idea is that in each iteration, we aggregate neighborhood scores for each node in all negative networks independently and then aggregate the scores of a node across all the networks. This aggregated score is used in the next iteration. 

In detail, we repeat the following three-step procedure until convergence (or until the maximum number of iterations is reached). 
In the first step, we initialize each node with an initial score in all the networks. Node deception prior scores are used for the first initialization. 
In the second step, in each negative network $E^{t-}$, we calculate the score $s^t(v)$ by aggregating node $v$'s outgoing neighbors' scores using the following equation: $s^t(v) = \beta \sum_{(v,u)\in E^{t-}} s^t(u)\cdot w_{v,u,t} +  (1-\beta)s^t(v)$. 
Here $\beta$ weighs the importance of a node's own deception score versus the aggregate of neighbors' scores.
Each neighbor $u$'s deception score is weighted by the weight of the outgoing edge from $v$ to $u$. 
In the third step, for each node $v$, we aggregate the scores of $v$ across all the negative networks to get $v$'s output score in current iteration, and normalize the scores. Averaging is used as the aggregation function; other aggregation functions, such as recency-weighted averaging, can be used instead, if desired.
The normalized output scores are used as the initial scores in the next iteration. 
After convergence, the final deceptiveness scores $s(V)$ are returned.

\textbf{Final classification.}
Finally, we train a binary classifier for predicting whether a participant is deceptive or not.
For a node $u$, the features we use are $u$'s final deceptiveness scores and $u$'s four behavior properties, namely the average fraction of speaking, average entropy of looking, average in-degree, and average in-degree while speaking.

{

\begin{algorithm}[!t]
\small
\KwIn{Negative dynamic interaction networks $[(V,E^{1-}),(V,E^{2-}),...,(V,E^{T-})]$, 
initial deceptive scores $s(V)$, $\beta$, convergence threshold $\tau$, maximum iteration number $M$}
\KwOut{Final deceptive scores $s(V)$}

$iter = 0, dif = 1$;

\While{ $dif > \tau$ and $iter < M$}{
$s^t(V) = s(V), \forall t=1 \ldots T$;

    \For{$t=1 \ldots T$}{
        \ForEach{$v \in V$}{
        $r^t(v) = \beta \sum_{(u,v)\in E^{t-}} s^t(u)\cdot w_{u,v,t} + \newline
        (1-\beta)s^t(v)$;
        }
    }
    
    $r(V) = \sum_t r^t(V) / T $;
    
    $r(V) = r(V) / \left\| r(V) \right\|_2$;
    
    $dif = \left\|(r(V) - s(V)) \right\|_2$; 
    
    $iter = iter + 1$;
    
    $s(V) = r(V)$
}
\Return $s(V)$
\caption{DeceptionRank on Negative Dynamic Interaction Networks}\label{alg:bp}
\end{algorithm}
}

\section{Experiments}
In this section, we compare the performance of DeceptionRank with state-of-the-art vision and graph embedding baselines. We show that DeceptionRank outperforms these baselines by at least 20.9\% in detecting deceivers. 

\subsection{Experiment Setup}
% \vspace{-0.3em}
The prediction task is: given a video segment of a game, predict the roles (deceiver or non-deceiver) of all participants in the game.

Since there are only 26 games in the dataset, we augment the dataset by segmenting long videos into several smaller videos. 
The roles of the players remain the same in the videos after segmenting. 
We split the games into 1 minute long video segments (we study the effect of segment length on prediction performance later).
This results in a dataset with 2781 data points. 

To ensure that there is no leakage of ground-truth labels, we split the dataset into training and test set according to games (not by video segments). Every player in our data participated in exactly one game --- so we never train on data of a player in one game and use that to predict whether he is deceptive or not in another.
We conduct all our experiments with 5-fold cross-validation, where all clips of a game belong to the same fold. 
\footnote{We do 5-fold cross validation rather than 10-fold because of the small number of games.}
This ensures that two segments of the same game can not appear in both training and test sets. 
Further, we split the data by participants as well, so a participant can only be either in training or test across all segments. 
In each fold, we set 60\% participants in the training set and the rest are in the test set (we ensure that at least one deceiver and non-deceiver are in training and test set).
Since the task is unbalanced, we report the AUROC, averaged over five folds.

\change{We do not provide any model, either our or the baselines, with the number of deceivers or non-deceivers in the game, so there is no leakage of label distribution. 
All the experiment setting are the same across all models to ensure fairness. 
}

\subsection{Baselines}
% \vspace{-0.3em}
We consider two sets of baselines: vision-based methods and graph embedding methods. All baselines are evaluated in the same setup and dataset as our model. 

\textbf{Computer vision baselines.}
We compare our method with five computer vision baselines with the same experimental setup as our method ~\cite{bai19autodecep,Baltrusaitis2018,Demyanov2015,bai19autodecep,wu2018deception}.
These methods used features extracted from the video, including facial emotion, head and eye movement, facial action units, and time-aggregated features as described below.

\change{\cite{Demyanov2015} extracts the averaged facial action units (FAUs) features over time. \cite{Baltrusaitis2018} computes eye movements from the estimated eye ball positions, and uses the movement distributions over time as features. \cite{wu2018deception} extracts the individual dense trajectory features from videos, MFCC features from audio, micro-expression features and text features from transcripts, and uses an ensemble method called late fusion to come up with a joint prediction. Since our dataset doesn't have transcripts and annotated micro-expressions, we remove the text features and replace micro-expressions by FAU features \cite{Demyanov2015}. 
Lastly, we extract the histograms of emotion features and LiarRank features proposed by \cite{bai19autodecep} as other two baselines, where LiarRank captures group information by ranking the feature values in each group as meta-features. 

Note that all these methods make predictions for each player individually, without considering interactions between players. 
Specifically, in these methods, we extract the feature values for each individual player and use them as input to train a binary classifier. 
All these baseline features are trained with Logistic Regression, Random Forest, Linear SVM and Navie Bayes. We report the best AUROC among these classifiers in Table~\ref{tab:results}.}

\textbf{Graph embedding baselines.}
Here we compare our method with dynamic graph embedding based methods. \change{Dynamic graph embedding models have shown incredible success in making predictions for large-scale social networks. }
In particular, we compare with temporal graph convolution networks (TGCN)~\cite{liu2019characterizing} on the look-at network, speak-to network, and listen-to networks.
\change{TGCN model combines graph convolution network with LSTM.
Given a sequence of networks and the ground-truth training node labels, TGCN trains two-layer GCN models on individual networks. All individual networks share the same GCN parameters. 
Mean pooling is used to aggregate neighborhood node information in the GCN. 
The sequence of output scores per node corresponding to the sequence of graphs is fed as input into an LSTM. 
All nodes share the same LSTM parameters. 
The final output of the LSTM is used to predict the training nodes' ground-truth label. 
The model is trained in an end-to-end manner, where the GCN and LSTM model parameters are trained to accurately predict the node labels.} 
We experimented with other variants of temporal graph models~\cite{zhou2018dynamic,goyal2018dyngem}, which gave similar performance. 

\textbf{Ensemble baseline.} We create an ensemble baseline model to combine the strengths of all the baselines. For each node, we concatenate its baseline scores from all the vision and graph embedding classifiers described above. This generates one feature vector per node, which is used as the node's input to a Logistic Regression classifier to make the prediction.

\begin{table}[t]
\centering
\resizebox{\columnwidth}{!}{
    
    \begin{tabular}{|c|c|c|}
    \hline
        \textbf{Method} & \textbf{Performance} & \textbf{\% Improvement}\\
        & & \textbf{Over Baseline}\\
        \hline
        
        \multicolumn{3}{|c|}{Computer Vision Baselines} \\\hline
        Emotions~\cite{bai19autodecep} & 0.538 & 39.9\%\\
        Movements~\cite{Baltrusaitis2018} & 0.549 & 37.2\%\\
        FAUs~\cite{Demyanov2015} & 0.569 & 32.3\%\\
        LiarRank~\cite{bai19autodecep} & 0.590 & 27.6\%\\
        Late fusion~\cite{wu2018deception} & 0.594 & 26.7\%\\\hline
        
        \multicolumn{3}{|c|}{Graph Embedding Baselines} \\\hline
        TGCN on Look-At~\cite{liu2019characterizing} & 0.550 & 36.9\% \\
        TGCN on Speak-To~\cite{liu2019characterizing} & 0.538 & 39.9\%\\
        TGCN on Listen-To~\cite{liu2019characterizing} & 0.541 & 39.2\%\\\hline
        
        \multicolumn{3}{|c|}{Ensemble Baseline} \\\hline
        Combining all the above features & 0.623 & 20.9\% \\\hline
        
        \multicolumn{3}{|c|}{Proposed Method} \\\hline
        \textbf{DeceptionRank} & \textbf{0.753} & - \\\hline
    \end{tabular}
    }
    
    \caption{
    Our proposed method DeceptionRank outperforms state-of-the-art vision, graph embedding, and ensemble baselines in the task of predicting deceivers from 1 minute clips of videos. 
    DeceptionRank outperforms all baselines by at least 20.9\% AUROC in prediction performance.
    \label{tab:results}
    }
\end{table}

\subsection{Predictive Performance of DeceptionRank}
\label{sec:performance}
Here we compare the  performance of DeceptionRank with the baselines. Table~\ref{tab:results} shows the cross-validation performance (in terms of AUROC) of all methods on 1 minute segments. We report the performance as average AUROC scores and their 95\% bootstrapped confidence interval.

First, we observe that DeceptionRank significantly outperforms all other methods, by at least 20.9\%. DeceptionRank has an AUROC of 0.753 with its 95\% confidence interval ranging from 0.721 to 0.789. 
Second, among baselines, the ensemble performs the best. We note that when used individually, vision-based baselines outperform baselines that use graph embeddings. We attribute this difference in performance to the small size of the dataset, leading to lower performance of deep-learning based GCN methods. 
Finally, LiarRank and Late Fusion outperform the other baselines. 
This is likely due to the fact that LiarRank was designed to identify deceivers in networks, while late fusion combines audio and transcripts with visual features. However, we remind readers that DeceptionRank generates the best performance compared to all baselines, suggesting that FFDINs and Negative Interaction Networks, together with the DeceptionRank algorithm generate excellent value in terms of performance.

\begin{figure}
\centering
        \includegraphics[width=0.9\columnwidth]{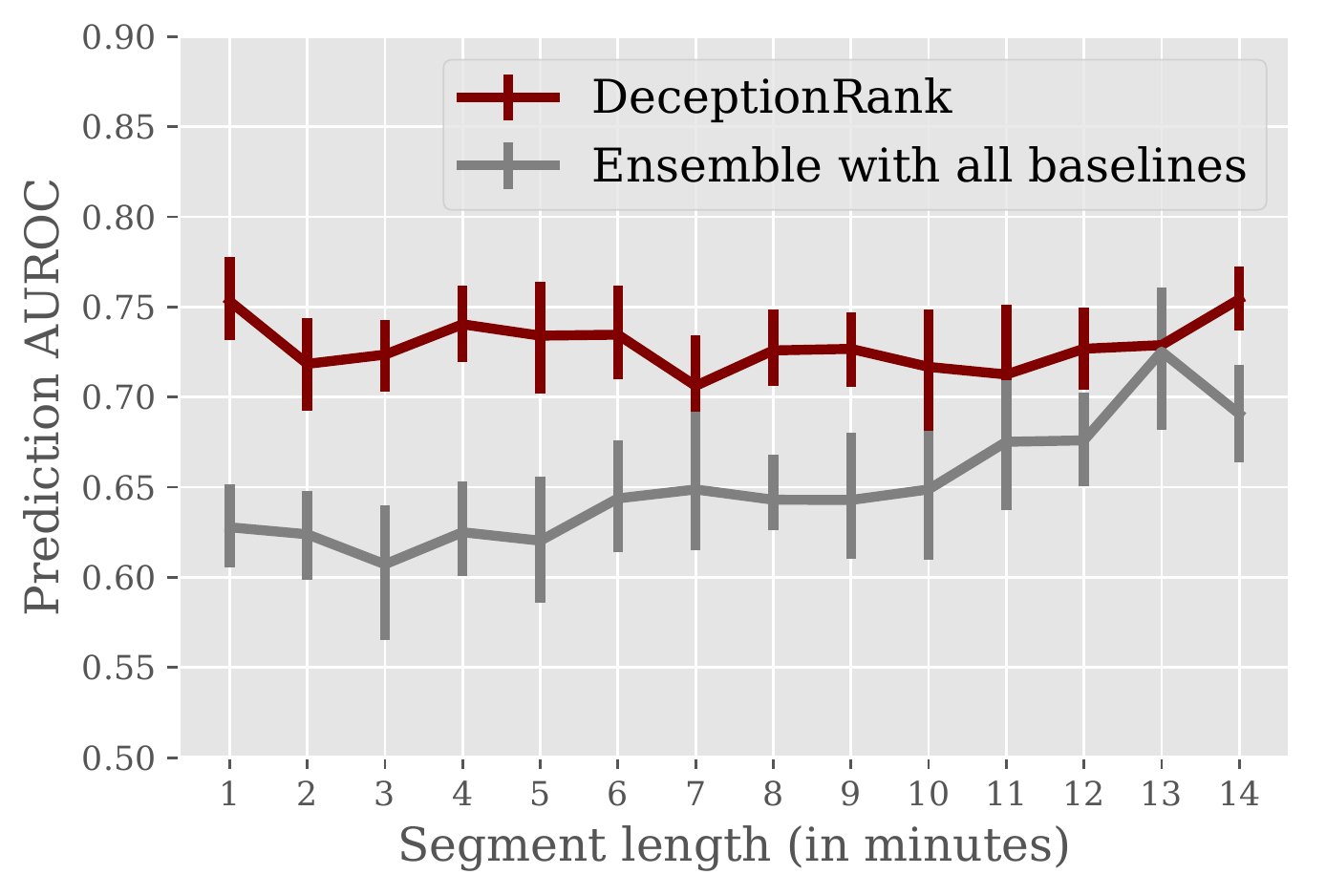} 
    % \vspace{-1em}
    \caption{\textit{DeceptionRank performs better than all features across all segment length.} 
    DeceptionRank has stable performance for all segment lengths too. 
    \label{fig:duration-impact}}
\end{figure}

\subsection{Performance vs. Segment Length}\label{sec:experiment-length}

The preceding experiment shows the performance of both DeceptionRank and the baselines using data from segments that are 1 minute long. 
We now study the impact of the input segment length on the  performance of predictors.
We vary segment length from 1 minute to 14 minutes.
For each segment length, we randomly sample 100 segments from each game. 
As before, we follow a 5-fold cross-validation setting. 
Finally, we compare the average AUROC and report the 95\% confidence interval of performance.

As the ensemble baseline performed the best among all baselines as shown in the previous experiment, we compare DeceptionRank with this ensemble baseline model. 
Other individual vision and graph embedding baselines have lower performances. 

Figure~\ref{fig:duration-impact} shows the results of varying the segment length. 
We show that DeceptionRank outperforms the best baseline for all segment lengths considered. The margin is large when the segment lengths considered are small --- once the segment lengths are over 10 minutes long, the performance of DeceptionRank and the baselines is similar. It is important to note
that DeceptionRank's performance is stable across segment lengths, while the baselines have diminished performance when the segments are short. 
These findings illustrate the robustness of our model with respect to the input durations.

\change{

\begin{figure}
\centering
        \includegraphics[width=0.9\columnwidth]{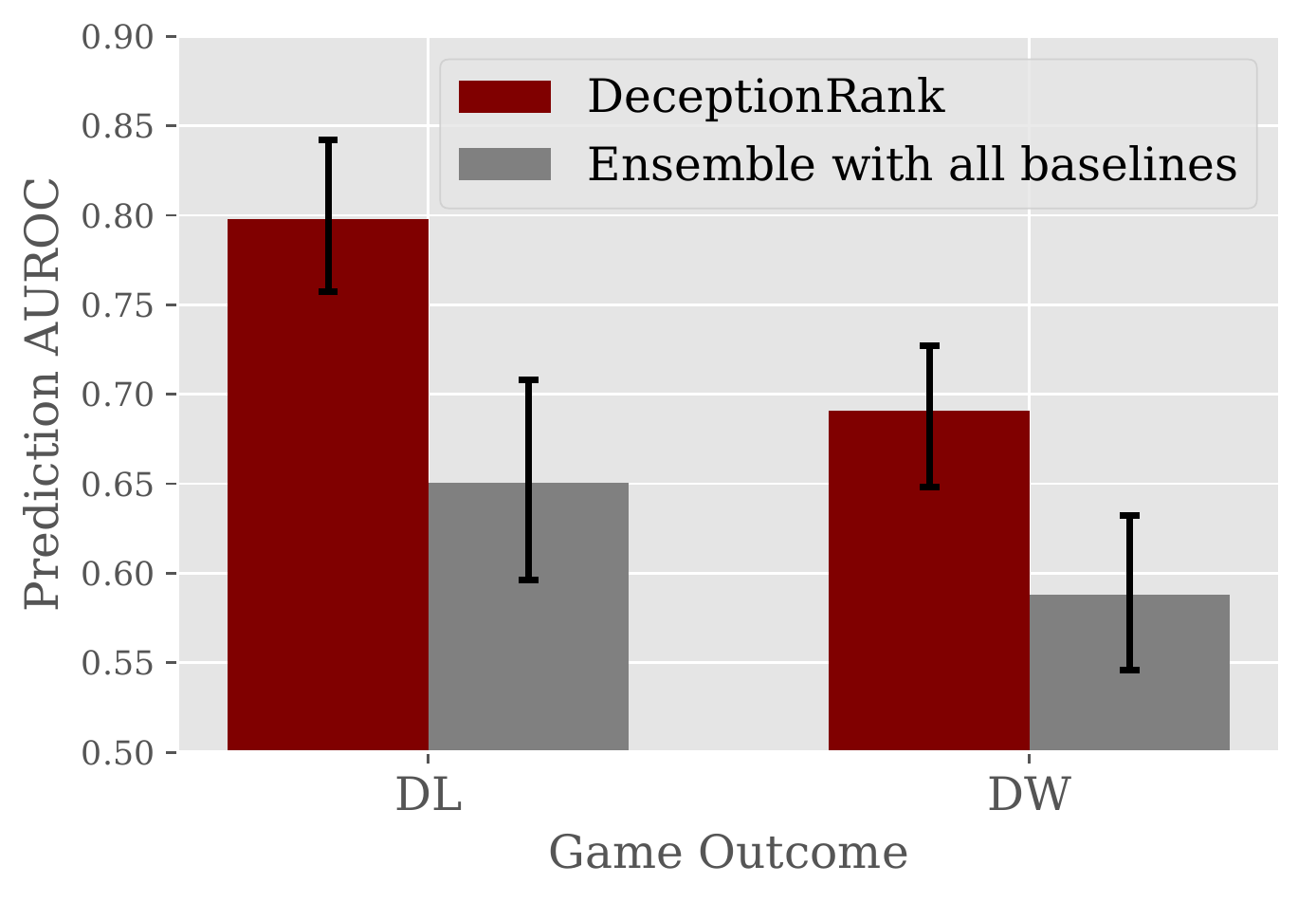} 
    % \vspace{-1em}
    \caption{\change{\textit{DeceptionRank performs better than all features across DL and DW games.}
    Model performances are better in DL games compared to DW games. }
    \label{fig:gametype-impact}}
\end{figure}

\subsection{Performance in DL vs. DW Games} 
Here we compare the performance of the models according to the game outcome. Specifically, we compare model performance for DL games vs. DW games. Recall that 14 out of 26 games (= 54\%) were won by deceivers. 
Since our analysis in the previous sections show that the deceivers differ significantly, we evaluate if that affects the models' performance across these games. 
As the ensemble baseline performed the best among all baselines, we compare DeceptionRank against the ensemble model.

We follow the default setting as outlined earlier for the experiments. We evaluate the models on 1 minute segments. We report the performance as average AUROC scores and their 95\% bootstrapped confidence interval. We randomly sample 100 segments from each game. As before, we follow a 5-fold cross-validation setting.

When evaluating the model performance for DL games, we consider only DL games, i.e., we train and test on DL games only. Similarly, only DW games are considered when evaluating model performance for DW games.

Figure~\ref{fig:gametype-impact} shows the results of according to the game outcome. 
First, we see that regardless of the game outcome, DeceptionRank performs better than the ensemble model.
Second, we note that both models perform better in DL games compared to their performance in the DW games. 
The explanation for this observation is that the behavior of deceivers in DL games is significantly different from that of non-deceivers. This makes it easier for the machine learning algorithms to distinguish between them. On the other hand, deceivers and non-deceivers have similar behavior in DW games, which makes it comparatively harder for the algorithms to identify them. 
These findings show the robustness of our model across game outcomes. 
}

In summary, the results in this section show that  DeceptionRank  outperforms the baseline methods by at least 20.9\% in identifying deceivers in groups.
Deceivers can be identified effectively even with short segments and better in DL games.

\section{Related Work}
We now summarize related work that was not already discussed earlier.

\textbf{Face-to-face deception.} 
There has been much research on predicting whether an individual is deceptive from facial and body cues ~\cite{ding2019face,randhavane2019liar,wang2020attention} with extensions that also include audio and linguistic cues ~\cite{gogate2017deep,wu2018deception}. \citet{ding2019face} proposed a deep CNN model to fuse face and body cues together which can be trained with limited data via meta-learning. \citet{randhavane2019liar} focused on deception prediction by feeding dynamic 3D gestures to an LSTM. They also identified deceptive behaviors (e.g. looking around, hands in pockets). \change{~\citet{wang2020attention} proposed the attention mechanism with 3D CNN to identify individual deceptive facial patterns.} ~\citet{wu2018deception} combined visual cues (e.g. micro-expression and video trajectory features) with voice and transcript data using a highly effective late fusion mechanism. 
However, there is limited work on predicting deception in multi-person face-to-face interaction settings. ~\citet{Chittaranjan2010} pioneered face to face deception detection from conversation cues (e.g. speaking turns and number of interruptions). 
\change{\citet{pak2013eye} used eye gaze to detect deception,} \citet{sapru2015automatic} used facial action units as features, and ~\citet{bai19autodecep} proposed the notion of LiarRank within groups to better capture group-based deception. 

However, most of these techniques do not work on group deception and do not consider inter-personal interactions in order to predict deception. We are the first to do so and to show that the Negative Dynamic Interaction Networks we propose are highly effective at detecting deception. This paper does not focus on computer vision --- rather it builds on the technique proposed by \citet{bai19vfoa} to extract the visual focus of attention of people --- from this, we build NDINs for deception detection.

\textbf{Deception detection on social media using social and interaction networks.} 
Extensive research has been done to detect deception using interaction and social networks. Many of these works have focused on web and social media domains. 
These include methods to detect fake news~\cite{liu2018early,horne2017just}, rumors~\cite{zeng2016unconfirmed,li2016user}, fake reviews \cite{mukherjee2013yelp,li2015analyzing}, spammers ~\cite{wu2017adaptive}, and coordinated activity~\cite{kumar2017army,subrahmanian2016darpa}. \citet{liu2018early} focused on the early detection of fake news on social media by modeling information spread on the network. % the paths along which news spreads along with the characteristics of related users using both CNN and LSTM models. 
\change{\citet{kumar2018rev2,hooi2016birdnest,rayana2015collective} leveraged the reviewer-product network to identify fake reviews and fraudulent reviewers in e-commerce platforms. }
\citet{wu2017adaptive} used sparse learning to detect spammers as a social community from both online social interactions and TF-IDF content features. 
\citet{kumar2017army} analyzed the behaviors of sockpuppets (users with multiple accounts to manipulate public opinions) in social media through multiple perspectives including their social networks, posting patterns, posting contents etc.  
\citet{subrahmanian2016darpa} developed a mix of language, network, and neighborhood features in order to identify both influence bots and botnets as part of a DARPA challenge.

\change{Though these papers use the concept of networks to study deception, they have not been applied to the video-based discussion settings. Compared to social networks, FFDINs extracted from videos are very different due to two important reasons. First, the edges in FFDINs represent instantaneous verbal and non-verbal interactions, and thus, the edges are highly dynamic. On the other hand, edges in social networks are relatively long-term and stable. 
Second, FFDINs have very few number of nodes and edges, while social networks have millions of nodes and edges. Both these major differences call for new methods that can work on small-scale but highly dynamic networks. Our DeceptionRank method bridges this gap. 
}

\change{
\textbf{Deception detection in games.}
The work most related to us are deception analysis from online chat-based mafia games~\cite{pak2015temporal,yu2015detecting}. \citet{pak2015temporal} builds a reply network over time, and hypothesizes several deceptive patterns such as centrality and nodes similarity. They conduct statistical analysis of the relationship to deception, but don't predict deceivers. \citet{yu2015detecting} builds a rule-based attitude network from the chat logs, and clusters nodes into subgroups 
of deceivers and non-deceivers. The clustering quality is measured by purity and entropy. Their results highly depend on the quality of chat logs and specific rules, which limit the application scope. Moreover, neither of these directly predicts the role for each player as we do.
\citet{niculae2015linguistic,azaria2015agent} studied the conversational properties patterns in a games to detect deception.
However, none of the above works have studied deception in face-to-face video communication, which is the gap we bridge. 
}

\section{Discussion and Conclusion}
To the best of our knowledge, this is the first paper to use network analysis methods in order to predict who is being deceptive in a video-based group interaction setting.  

Using a dataset based on the well-known Resistance game, we propose the concepts of a Face-to-Face Dynamic Interaction Network (FFDIN) and a Negative Dynamic Interaction Network (NDIN). We propose the DeceptionRank algorithm, and show that DeceptionRank beats several baselines including ones based on computer vision and graph embedding in detecting deceivers. 

\change{
\textbf{Relevance to the web and social media community.}
Our research sheds light on group deception in video-based conversations. 
While our work focuses on conversations in a social game, such conversations are commonplace in everyday communications via video call apps such as Microsoft Teams, Google Meet, Facebook Messenger, Zoom and Skype, and form an integral part of social media platforms like Facebook, SnapChat, and WhatsApp. 
The inputs are look-at, speak-to and listen-to interactions between people, which can be extracted from these videos from web and social media platforms. 
The proposed methods of Negative Dynamic Interaction Network and DeceptionRank can then be applied to identify deceivers in those videos.
Since there are no such web and social media datasets with ground-truth of deception, we leave the experimental evaluation of our methods on web and social media video-based deception for future work. 
Techniques in our work have the potential to improve the safety and integrity of social media and web-based communication platforms. 

\textbf{Use of one dataset.}
Currently, there are no other datasets of face-to-face video conversations with ground truth of group deception on which we can test our model. This is because creating such a dataset is an extremely difficult and time-consuming effort. This is highlighted in the paper \cite{bai19vfoa} from which we have derived our dataset --- it took the authors about 18 months to collect and process the 26 videos. Testing generalizing capability of our method beyond the current dataset will require creation of new datasets, a huge task which can be conducted in the future. 

The only other comparable video dataset is given by \cite{perez2015deception} which contains 56 people and spans 57 minutes, but each video only has one person. There is no group interaction or deception. So, we can not use it for our task. By comparison, the dataset we use contains 185 participants and spans 1000 minutes. Our dataset is significantly larger.

\textbf{Generalizing beyond a game.}
Although the setting we study in this work is in a social game setting, the discussions are free-form and the participants can deceive others as they want. No instructions or training was provided to them about how to deceive. Thus, the findings in this paper should represent the general properties of how deceivers operate in groups. Importantly, the Negative Dynamic Interaction Network and DeceptionRank method are general and can be applied to any setting involving interactions between groups of people. 
}

\textbf{Future work.}
Future work can expand this study to multiple games and settings such as sales meetings, business negotiations, job interviews, and more.
The methods of Negative Dynamic Interaction Networks and DeceptionRank can be tested for deception detection in other datasets, including social network datasets. 
Finally, our methods can be used to study other social affects, such as leadership, trust, liking, and dominance.

\section{Acknowledgements}
% This work was funded in part by ARO Grant W911NF1610342. 
We gratefully acknowledge the support of 
NSF under Nos. OAC-1835598 (CINES), OAC-1934578 (HDR), CCF-1918940 (Expeditions), IIS-2030477 (RAPID), IIS-2027689 (RAPID);
DARPA under No. N660011924033 (MCS);
ARO under Nos. W911NF-16-1-0342 (MURI), W911NF-16-1-0171 (DURIP);
Stanford Data Science Initiative, 
Wu Tsai Neurosciences Institute,
Chan Zuckerberg Biohub,
Amazon, JPMorgan Chase, Docomo, Hitachi, JD.com, KDDI, NVIDIA, Dell, Toshiba, UnitedHealth Group, Adobe, Facebook, Microsoft, and IDEaS Insitute. 
J. L. is a Chan Zuckerberg Biohub investigator.

\bibliographystyle{aaai}
\bibliography{bib/refs.bib}

\begin{thebibliography}{48}
\providecommand{\natexlab}[1]{#1}
\providecommand{\url}[1]{\texttt{#1}}
\providecommand{\urlprefix}{URL }
\expandafter\ifx\csname urlstyle\endcsname\relax
  \providecommand{\doi}[1]{doi:\discretionary{}{}{}#1}\else
  \providecommand{\doi}{doi:\discretionary{}{}{}\begingroup
  \urlstyle{rm}\Url}\fi

\bibitem[{Addawood et~al.(2019)Addawood, Badawy, Lerman, and
  Ferrara}]{addawood2019linguistic}
Addawood, A.; Badawy, A.; Lerman, K.; and Ferrara, E. 2019.
\newblock Linguistic cues to deception: Identifying political trolls on social
  media.
\newblock In \emph{International AAAI Conference on Web and Social Media},
  volume~13, 15--25.

\bibitem[{Azaria, Richardson, and Kraus(2015)}]{azaria2015agent}
Azaria, A.; Richardson, A.; and Kraus, S. 2015.
\newblock An agent for deception detection in discussion based environments.
\newblock In \emph{Proceedings of the 18th ACM Conference on Computer Supported
  Cooperative Work \& Social Computing}, 218--227.

\bibitem[{Baccarani and Bonfanti(2015)}]{baccarani2015effective}
Baccarani, C.; and Bonfanti, A. 2015.
\newblock Effective public speaking: a conceptual framework in the
  corporate-communication field.
\newblock \emph{Corporate Communications: An International Journal} .

\bibitem[{{Bai} et~al.(2019){Bai}, {Bolonkin}, {Burgoon}, {Chen}, {Dunbar},
  {Singh}, {Subrahmanian}, and {Wu}}]{bai19autodecep}
{Bai}, C.; {Bolonkin}, M.; {Burgoon}, J.; {Chen}, C.; {Dunbar}, N.; {Singh},
  B.; {Subrahmanian}, V.~S.; and {Wu}, Z. 2019.
\newblock Automatic long-term deception detection in group interaction videos.
\newblock In \emph{IEEE International Conference on Multimedia and Expo}.

\bibitem[{Bai et~al.(2019)Bai, Kumar, Leskovec, Metzger, Nunamaker, and
  Subrahmanian}]{bai19vfoa}
Bai, C.; Kumar, S.; Leskovec, J.; Metzger, M.; Nunamaker, J.~F.; and
  Subrahmanian, V. 2019.
\newblock Predicting the visual focus of attention in multi-person discussion
  videos.
\newblock In \emph{International Joint Conference on Artificial Intelligence}.

\bibitem[{Baltrusaitis et~al.(2018)Baltrusaitis, Zadeh, Lim, and
  Morency}]{Baltrusaitis2018}
Baltrusaitis, T.; Zadeh, A.~B.; Lim, Y.~C.; and Morency, L.-P. 2018.
\newblock OpenFace 2.0: Facial behavior analysis toolkit.
\newblock In \emph{IEEE International Conference on Automatic Face and Gesture
  Recognition}.

\bibitem[{Beslin and Reddin(2004)}]{beslin2004leaders}
Beslin, R.; and Reddin, C. 2004.
\newblock How leaders can communicate to build trust.
\newblock \emph{Ivey business journal} 69(2): 1--6.

\bibitem[{Chittaranjan and Hung(2010)}]{Chittaranjan2010}
Chittaranjan, G.; and Hung, H. 2010.
\newblock Are you A werewolf? Detecting deceptive roles and outcomes in a
  conversational role-playing game.
\newblock In \emph{international Conference on Acoustics, Speech, \& Signal
  Processing}.

\bibitem[{Demyanov et~al.(2015)Demyanov, Bailey, Ramamohanarao, and
  Leckie}]{Demyanov2015}
Demyanov, S.; Bailey, J.; Ramamohanarao, K.; and Leckie, C. 2015.
\newblock Detection of deception in the mafia party game.
\newblock In \emph{ACM International Conference on Multimodal Interaction}.

\bibitem[{Derber(2000)}]{derber2000pursuit}
Derber, C. 2000.
\newblock \emph{The pursuit of attention: Power and ego in everyday life}.
\newblock Oxford University Press.

\bibitem[{Ding et~al.(2019)Ding, Zhao, Lu, Xiang, and Wen}]{ding2019face}
Ding, M.; Zhao, A.; Lu, Z.; Xiang, T.; and Wen, J.-R. 2019.
\newblock Face-focused cross-stream network for deception detection in videos.
\newblock In \emph{IEEE Conference on Computer Vision and Pattern Recognition}.

\bibitem[{Dinges et~al.(2005)Dinges, Rider, Dorrian, McGlinchey, Rogers,
  Cizman, Goldenstein, Vogler, Venkataraman, and Metaxas}]{dinges2005optical}
Dinges, D.~F.; Rider, R.~L.; Dorrian, J.; McGlinchey, E.~L.; Rogers, N.~L.;
  Cizman, Z.; Goldenstein, S.~K.; Vogler, C.; Venkataraman, S.; and Metaxas,
  D.~N. 2005.
\newblock Optical computer recognition of facial expressions associated with
  stress induced by performance demands.
\newblock \emph{Aviation, space, and environmental medicine} 76(6): B172--B182.

\bibitem[{Driskell, Salas, and Driskell(2012)}]{driskell2012social}
Driskell, J.~E.; Salas, E.; and Driskell, T. 2012.
\newblock Social indicators of deception.
\newblock \emph{Human Factors} 54(4): 577--588.

\bibitem[{Ellsberg(2010)}]{ellsberg2010power}
Ellsberg, M. 2010.
\newblock \emph{The power of eye contact: Your secret for success in business,
  love, and life}.
\newblock Harper Paperbacks.

\bibitem[{Gogate, Adeel, and Hussain(2017)}]{gogate2017deep}
Gogate, M.; Adeel, A.; and Hussain, A. 2017.
\newblock Deep learning driven multimodal fusion for automated deception
  detection.
\newblock In \emph{IEEE Symposium Series on Computational Intelligence}.

\bibitem[{Goyal et~al.(2018)Goyal, Kamra, He, and Liu}]{goyal2018dyngem}
Goyal, P.; Kamra, N.; He, X.; and Liu, Y. 2018.
\newblock DynGEM: Deep embedding method for dynamic graphs.
\newblock \emph{arXiv preprint arXiv:1805.11273} .

\bibitem[{Hooi et~al.(2016)Hooi, Shah, Beutel, G{\"u}nnemann, Akoglu, Kumar,
  Makhija, and Faloutsos}]{hooi2016birdnest}
Hooi, B.; Shah, N.; Beutel, A.; G{\"u}nnemann, S.; Akoglu, L.; Kumar, M.;
  Makhija, D.; and Faloutsos, C. 2016.
\newblock Birdnest: Bayesian inference for ratings-fraud detection.
\newblock In \emph{Proceedings of the 2016 SIAM International Conference on
  Data Mining}, 495--503. SIAM.

\bibitem[{Horne and Adali(2017)}]{horne2017just}
Horne, B.~D.; and Adali, S. 2017.
\newblock This just in: fake news packs a lot in title, uses simpler,
  repetitive content in text body, more similar to satire than real news.
\newblock In \emph{International AAAI Conference on Web and Social Media}.

\bibitem[{Keller et~al.(2017)Keller, Schoch, Stier, and
  Yang}]{keller2017manipulate}
Keller, F.~B.; Schoch, D.; Stier, S.; and Yang, J. 2017.
\newblock How to manipulate social media: Analyzing political astroturfing
  using ground truth data from South Korea.
\newblock In \emph{International AAAI Conference on Web and Social Media}.

\bibitem[{Kumar et~al.(2017)Kumar, Cheng, Leskovec, and
  Subrahmanian}]{kumar2017army}
Kumar, S.; Cheng, J.; Leskovec, J.; and Subrahmanian, V. 2017.
\newblock An army of me: Sockpuppets in online discussion communities.
\newblock In \emph{Proceedings of the International World Wide Web Conference}.

\bibitem[{Kumar et~al.(2018)Kumar, Hooi, Makhija, Kumar, Faloutsos, and
  Subrahmanian}]{kumar2018rev2}
Kumar, S.; Hooi, B.; Makhija, D.; Kumar, M.; Faloutsos, C.; and Subrahmanian,
  V. 2018.
\newblock Rev2: Fraudulent user prediction in rating platforms.
\newblock In \emph{Proceedings of the Eleventh ACM International Conference on
  Web Search and Data Mining}, 333--341.

\bibitem[{Kumar, Spezzano, and Subrahmanian(2015)}]{kumar2015vews}
Kumar, S.; Spezzano, F.; and Subrahmanian, V. 2015.
\newblock Vews: A wikipedia vandal early warning system.
\newblock In \emph{Proceedings of the 21th ACM SIGKDD international conference
  on knowledge discovery and data mining}, 607--616.

\bibitem[{Kumar, Zhang, and Leskovec(2019)}]{kumar2019predicting}
Kumar, S.; Zhang, X.; and Leskovec, J. 2019.
\newblock Predicting dynamic embedding trajectory in temporal interaction
  networks.
\newblock In \emph{Proceedings of the 25th ACM SIGKDD International Conference
  on Knowledge Discovery \& Data Mining}, 1269--1278.

\bibitem[{Laretzaki et~al.(2011)Laretzaki, Plainis, Vrettos, Chrisoulakis,
  Pallikaris, and Bitsios}]{laretzaki2011threat}
Laretzaki, G.; Plainis, S.; Vrettos, I.; Chrisoulakis, A.; Pallikaris, I.; and
  Bitsios, P. 2011.
\newblock Threat and trait anxiety affect stability of gaze fixation.
\newblock \emph{Biological psychology} 86(3): 330--336.

\bibitem[{Li et~al.(2015)Li, Chen, Mukherjee, Liu, and Shao}]{li2015analyzing}
Li, H.; Chen, Z.; Mukherjee, A.; Liu, B.; and Shao, J. 2015.
\newblock Analyzing and detecting opinion spam on a large-scale dataset via
  temporal and spatial patterns.
\newblock In \emph{International AAAI Conference on Web and Social Media}.

\bibitem[{Li et~al.(2016)Li, Liu, Fang, Nourbakhsh, and Shah}]{li2016user}
Li, Q.; Liu, X.; Fang, R.; Nourbakhsh, A.; and Shah, S. 2016.
\newblock User behaviors in newsworthy rumors: A case study of twitter.
\newblock In \emph{International AAAI Conference on Web and Social Media}.

\bibitem[{Liu et~al.(2019)Liu, Shi, Pierce, and Ren}]{liu2019characterizing}
Liu, Y.; Shi, X.; Pierce, L.; and Ren, X. 2019.
\newblock Characterizing and forecasting user engagement with in-app action
  graph: A case study of Snapchat.
\newblock In \emph{ACM SIGKDD International Conference on Knowledge Discovery
  and Data Mining}.

\bibitem[{Liu and Wu(2018)}]{liu2018early}
Liu, Y.; and Wu, Y.-F.~B. 2018.
\newblock Early detection of fake news on social media through propagation path
  classification with recurrent and convolutional networks.
\newblock In \emph{AAAI Conference on Artificial Intelligence}.

\bibitem[{Mukherjee et~al.(2013)Mukherjee, Venkataraman, Liu, and
  Glance}]{mukherjee2013yelp}
Mukherjee, A.; Venkataraman, V.; Liu, B.; and Glance, N. 2013.
\newblock What yelp fake review filter might be doing?
\newblock In \emph{International AAAI Conference on Weblogs and Social Media}.

\bibitem[{Niculae et~al.(2015)Niculae, Kumar, Boyd-Graber, and
  Danescu-Niculescu-Mizil}]{niculae2015linguistic}
Niculae, V.; Kumar, S.; Boyd-Graber, J.; and Danescu-Niculescu-Mizil, C. 2015.
\newblock Linguistic Harbingers of Betrayal: A Case Study on an Online Strategy
  Game.
\newblock In \emph{Proceedings of the 53rd Annual Meeting of the Association
  for Computational Linguistics}.

\bibitem[{Page et~al.(1999)Page, Brin, Motwani, and
  Winograd}]{page1999pagerank}
Page, L.; Brin, S.; Motwani, R.; and Winograd, T. 1999.
\newblock The PageRank citation ranking: Bringing order to the web.
\newblock Technical report, Stanford InfoLab.

\bibitem[{Pak and Zhou(2013)}]{pak2013eye}
Pak, J.; and Zhou, L. 2013.
\newblock Eye gazing behaviors in online deception.
\newblock In \emph{19th Americas Conference on Information Systems}.

\bibitem[{Pak and Zhou(2015)}]{pak2015temporal}
Pak, J.; and Zhou, L. 2015.
\newblock Temporal Patterns of Structural Deception Behavior in a Massively
  Multiplayer Online Game.
\newblock In \emph{2015 48th Hawaii International Conference on System
  Sciences}, 131--140. IEEE.

\bibitem[{P{\'e}rez-Rosas et~al.(2015)P{\'e}rez-Rosas, Abouelenien, Mihalcea,
  and Burzo}]{perez2015deception}
P{\'e}rez-Rosas, V.; Abouelenien, M.; Mihalcea, R.; and Burzo, M. 2015.
\newblock Deception detection using real-life trial data.
\newblock In \emph{Proceedings of the 2015 ACM on International Conference on
  Multimodal Interaction}, 59--66.

\bibitem[{Randhavane et~al.(2019)Randhavane, Bhattacharya, Kapsaskis, Gray,
  Bera, and Manocha}]{randhavane2019liar}
Randhavane, T.; Bhattacharya, U.; Kapsaskis, K.; Gray, K.; Bera, A.; and
  Manocha, D. 2019.
\newblock The Liar's Walk: Detecting Deception with Gait and Gesture.
\newblock \emph{arXiv preprint arXiv:1912.06874} .

\bibitem[{Rayana and Akoglu(2015)}]{rayana2015collective}
Rayana, S.; and Akoglu, L. 2015.
\newblock Collective opinion spam detection: Bridging review networks and
  metadata.
\newblock In \emph{Proceedings of the 21th ACM SIGKDD International Conference
  on Knowledge Discovery and Data Mining}, 985--994.

\bibitem[{Rayner(1998)}]{rayner1998eye}
Rayner, K. 1998.
\newblock Eye movements in reading and information processing: 20 years of
  research.
\newblock \emph{Psychological bulletin} 124(3): 372.

\bibitem[{Sapru and Bourlard(2015)}]{sapru2015automatic}
Sapru, A.; and Bourlard, H. 2015.
\newblock Automatic recognition of emergent social roles in small group
  interactions.
\newblock \emph{IEEE Transactions on Multimedia} 17(5): 746--760.

\bibitem[{Str{\"o}fer et~al.(2016)Str{\"o}fer, Ufkes, Noordzij, and
  Giebels}]{strofer2016catching}
Str{\"o}fer, S.; Ufkes, E.~G.; Noordzij, M.~L.; and Giebels, E. 2016.
\newblock Catching a deceiver in the act: Processes underlying deception in an
  interactive interview setting.
\newblock \emph{Applied psychophysiology and biofeedback} 41(3): 349--362.

\bibitem[{Subrahmanian et~al.(2016)Subrahmanian, Azaria, Durst, Kagan,
  Galstyan, Lerman, Zhu, Ferrara, Flammini, and
  Menczer}]{subrahmanian2016darpa}
Subrahmanian, V.; Azaria, A.; Durst, S.; Kagan, V.; Galstyan, A.; Lerman, K.;
  Zhu, L.; Ferrara, E.; Flammini, A.; and Menczer, F. 2016.
\newblock The DARPA Twitter bot challenge.
\newblock \emph{Computer} 49(6): 38--46.

\bibitem[{Vrij(2008)}]{vrij2008detecting}
Vrij, A. 2008.
\newblock \emph{Detecting lies and deceit: Pitfalls and opportunities}.
\newblock John Wiley \& Sons.

\bibitem[{Wang et~al.(2020)Wang, Bai, Bolonkin, Burgoon, Dunbar, Subrahmanian,
  and Metaxas}]{wang2020attention}
Wang, L.; Bai, C.; Bolonkin, M.; Burgoon, J.; Dunbar, N.; Subrahmanian, V.; and
  Metaxas, D.~N. 2020.
\newblock Attention-based Facial Behavior Analytics in Social Communication.
\newblock In \emph{30th British Machine Vision Conference, BMVC 2019}.

\bibitem[{Wiseman(2010)}]{wiseman201059}
Wiseman, R. 2010.
\newblock \emph{59 Seconds: Change Your Life in Under a Minute}.
\newblock Vintage.

\bibitem[{Wu et~al.(2017)Wu, Hu, Morstatter, and Liu}]{wu2017adaptive}
Wu, L.; Hu, X.; Morstatter, F.; and Liu, H. 2017.
\newblock Adaptive spammer detection with sparse group modeling.
\newblock In \emph{International AAAI Conference on Web and Social Media}.

\bibitem[{Wu et~al.(2018)Wu, Singh, Davis, and Subrahmanian}]{wu2018deception}
Wu, Z.; Singh, B.; Davis, L.~S.; and Subrahmanian, V. 2018.
\newblock Deception detection in videos.
\newblock In \emph{AAAI Conference on Artificial Intelligence}.

\bibitem[{Yu et~al.(2015)Yu, Tyshchuk, Ji, and Wallace}]{yu2015detecting}
Yu, D.; Tyshchuk, Y.; Ji, H.; and Wallace, W. 2015.
\newblock Detecting deceptive groups using conversations and network analysis.
\newblock In \emph{Proceedings of the 53rd Annual Meeting of the Association
  for Computational Linguistics}.

\bibitem[{Zeng, Starbird, and Spiro(2016)}]{zeng2016unconfirmed}
Zeng, L.; Starbird, K.; and Spiro, E.~S. 2016.
\newblock \# unconfirmed: Classifying rumor stance in crisis-related social
  media messages.
\newblock In \emph{International AAAI Conference on Web and Social Media}.

\bibitem[{Zhou et~al.(2018)Zhou, Yang, Ren, Wu, and Zhuang}]{zhou2018dynamic}
Zhou, L.; Yang, Y.; Ren, X.; Wu, F.; and Zhuang, Y. 2018.
\newblock Dynamic network embedding by modeling triadic closure process.
\newblock In \emph{AAAI Conference on Artificial Intelligence}.

\end{thebibliography}

\end{document}